\documentclass[journal=jacsat,manuscript=article]{achemso}

\usepackage[version=3]{mhchem} 
\usepackage[colorlinks=true,linkcolor=red]{hyperref}

\author{Jiayan Xu}
\affiliation{Department of Chemistry, Princeton University, Princeton, NJ, United States}
\author{Shreeja Das}
\affiliation{Shell India Markets Pvt. Ltd., Bengaluru, India}
\author{Amar Deep Pathak}
\affiliation{Shell India Markets Pvt. Ltd., Bengaluru, India}
\email{Amar-Deep.Pathak@shell.com}
\author{Abhirup Patra}
\affiliation{Shell International Exploration \& Production Inc., Houston, TX, United States}
\author{Sharan Shetty}
\affiliation{Shell India Markets Pvt. Ltd., Bengaluru, India}
\author{Detlef Hohl}
\affiliation{Shell International Exploration \& Production Inc., Houston, TX, United States}
\author{Roberto Car}
\affiliation{Department of Chemistry, Princeton University, Princeton, NJ, United States}
\email{rcar@princeton.edu}

\title{Dynamic Metal-Support Interaction Dictates Cu Nanoparticle Sintering on Al$_\mathbf{2}$O$_\mathbf{3}$ Surfaces}


\begin{document}

%

\newpage

\begin{abstract}
    Nanoparticle sintering remains a critical challenge in heterogeneous catalysis.
    In this work, we present a unified deep potential (DP) model for Cu nanoparticles on three Al$_2$O$_3$ surfaces ($\gamma$-Al$_2$O$_3$(100), $\gamma$-Al$_2$O$_3$(110), and $\alpha$-Al$_2$O$_3$(0001)).
    Using DP-accelerated simulations, we reveal striking facet-dependent nanoparticle stability and mobility patterns across the three surfaces. 
    The nanoparticles diffuse several times faster on $\alpha$-Al$_2$O$_3$(0001) than on $\gamma$-Al$_2$O$_3$(100) at 800 K while expected to be more sluggish based on their larger binding energy at 0 K.
    Diffusion is facilitated by dynamic metal-support interaction (MSI), where the Al atoms switch out of the surface plane to optimize contact with the nanoparticle and relax back to the plane as the nanoparticle moves away. In contrast, the MSI on $\gamma$-Al$_2$O$_3$(100) and on $\gamma$-Al$_2$O$_3$(110) is dominated by more stable and directional Cu-O bonds, consistent with the limited diffusion observed on these surfaces.
    Our extended long-time MD simulations provide quantitative insights into the sintering processes, showing that the dispersity of nanoparticles (the initial inter-nanoparticle distance) strongly influences coalescence driven by nanoparticle diffusion.
    We observed that the coalescence of Cu$_{13}$ nanoparticles on $\alpha$-Al$_2$O$_3$(0001) can occur in a short time (10 ns) at 800 K even with an initial inter-nanoparticle distance increased to 30 \r{A}, while the coalescence on $\gamma$-Al$_2$O$_3$(100) is inhibited significantly by increasing the initial inter-nanoparticle distance from 15 \r{A} to 30 \r{A}.
    These findings demonstrate that the dynamics of the supporting surface is crucial to understanding the sintering mechanism and offer guidance for designing sinter-resistant catalysts by engineering the support morphology.
\end{abstract}
\newpage

\section{Introduction}
Supported metal nanoparticles, while the cornerstone of modern heterogeneous catalysis, face critical challenges in maintaining steady catalytic performance in activity and selectivity for a long period.\cite{Li2020ChemRev}
The overall catalytic performance of these nanoparticles is often compromised by deactivation under \textit{operando} conditions, which is an inherent part of most catalytic processes, reduces the lifetime of the catalyst and severely affects industrial production.
Sintering is one of the main factors in the deactivation of supported nanoparticle catalysts and is known to be governed by the metal-support interaction (MSI) that changes throughout the operational time depending on reaction conditions.\cite{Tao2008Science,Yoshida2012Science,Li2015NatCommun,Foucher2023CatalToday}
A consensus from several studies is that sintering proceeds via two pathways: Ostwald ripening (involving the migration of atomic species from smaller to larger particles),  and coalescence (where entire particles move and merge).\cite{Dai2018ChemSocRev, Campbell2013AccChemRes}
Hu and Li in a recent work proposed the Sabatier principle for the sintering process based on a quantitative study of MSI.
They found a linear relationship between the nanoparticle-support adhesion energy and the onset temperature of sintering, focusing primarily on fully relaxed, static idealized surfaces.
They concluded that a stronger MSI leads to Ostwald ripening, while a weaker MSI is responsible for coalescence on homogeneous supports\cite{Hu2021Science}.
However, MSI is a complex, multifaceted phenomenon that encompasses not only binding and adhesion energies at zero temperature but also the dynamic interplay between the nanoparticle and the support, which may not be fully captured by static models.
It remains elusive to predict how realistic conditions determine catalyst stability when the complex interplay between nanoparticle size, support structure, and temperature is taken into account.

Cu-supported catalysts find extensive applications in the chemical industry,\cite{Gawande2016ChemRev} particularly in CO$_2$ hydrogenation and methanol synthesis.\cite{Behrens2012Science,Kuld2016Science,Kattel2017Science,Lam2019Angew} 
These catalytic reactions are often structure sensitive, dictated by the Cu nanoparticle size and shape.\cite{Berg2016NatComm}
Hence, the choice of support is crucial to achieving optimal MSI in Cu-catalyzed reactions, ensuring a stable size distribution of Cu nanoparticles that maintain activity over an extended period of operations. Al$_2$O$_3$ is one of the commonly used supports. In particular, $\gamma$-Al$_2$O$_3$ is widely preferred over $\alpha$-Al$_2$O$_3$ for Cu nanoparticles due to its high specific surface area and porosity resulting in higher nanoparticle dispersion.\cite{Argyle2015Catalysts} However, thermal sintering and deactivation of supported nanoparticles is unavoidable under experimental conditions and depends on the surface termination.\cite{Kurtz2003CatalLett} An experimental study showed that the sintering on Al-terminated $\alpha$-Al$_2$O$_3$(0001) occurs when a high coverage of Cu nanoparticles is deposited.\cite{Jensen2008JPCC}
From an atomic perspective, unravelling the morphological changes of the Cu nanoparticles and their dynamical interaction with different Al$_2$O$_3$ surfaces is of paramount importance for designing robust supported Cu nanoparticle catalysts with enhanced lifetime and activity.

Computational modeling has emerged as a powerful tool for gaining atomic insights into supported nanoparticle catalysts.
Although density functional theory (DFT) has been instrumental in understanding supported nanoparticle catalysts, its high computational cost limits the studies to small, often unrealistic model systems, containing up to a few dozen metal atoms.\cite{Liu2015JACS,Sun2018JACS,Zhou2023MolCatal}
Moreover, supported nanoparticles are inherently dynamic under reaction conditions, so it is important to sample a set of diverse thermally accessible configurations that can influence catalytic properties. Capturing this structural fluxionality is crucial for the accurate modeling of catalytic behavior, but presents a significant computational challenge for conventional DFT approaches.\cite{Zhang2020AccChemRes}
Recent advances in machine learning interatomic potentials (MLIPs) have opened up exciting possibilities for modeling complex catalyst systems with \textit{ab initio} accuracy at a fraction of computational cost.\cite{Xu2021PCCP,Cheng2024PrecisChem}
By training on a diverse dataset of DFT calculations, MLIPs can learn the underlying potential energy surface and enable molecular dynamics (MD) simulations of realistic supported nanoparticle models over extended time and length scales.
In fact, some recent studies have made efforts to train MLIPs for a few specific combinations of metal nanoparticles and supports\cite{Liu2022JACS,Xuan2024NanoLett}
However, training robust and transferable MLIPs for supported nanoparticle systems is a nontrivial task, especially for various combinations of nanoparticle sizes and supports, requiring careful data set curation and efficient active learning strategies to sample relevant regions in the configuration space.

Here, we present a systematic workflow to train a unified deep potential (DP) model for Cu nanoparticles supported on different surface terminations of $\gamma$- and $\alpha$-Al$_2$O$_3$.
Our training data set is constructed by combining global optimization (GO) techniques with MD sampling that capture finite-temperature effects.
The resulting DP model is capable of accurately describing supported Cu nanoparticles of different sizes at a finite temperature.
The model is validated against DFT-optimized global minimum configurations of supported Cu nanoparticles up to Cu$_{21}$ and against \textit{ab initio} MD (AIMD) simulations for three selected nanoparticle sizes within a short time window (20 ps) on $\gamma$-Al$_2$O$_3$(100), $\gamma$-Al$_2$O$_3$(110) and $\alpha$-Al$_2$O$_3$(0001). 
This validation allows us to explore the thermodynamic stability and sintering behavior at the atomic level.
We found that Cu nanoparticles tend to diffuse faster on $\alpha$-Al$_2$O$_3$(0001) than on $\gamma$-Al$_2$O$_3$(100) despite their stronger bindings in configurations that minimize the potential energy. 
The unexpected mobility on $\alpha$-Al$_2$O$_3$(0001) originates from the dynamics of the supporting surface driven by the formation of metallic bonds between the Cu and Al atoms.
In the presence of a nanoparticle the closest Al atoms relax quickly out of the surface plane to optimize the contact with the nanoparticle and relax back to the surface plane as the nanoparticle diffuses away, which we call the dynamic MSI.
Further MD simulations of nine supported Cu$_{13}$ nanoparticles reveal that coalescence can be observed on $\alpha$-Al$_2$O$_3$(0001) in 10 ns at 800 K even with an initial distance between nanoparticles increased to approximately 30 \r{A} while the coalescence on $\gamma$-Al$_2$O$_3$(100) is significantly inhibited by increasing the initial distance between nanoparticles from 15 \r{A} to 30 \r{A}.
Besides, nanoparticles on $\gamma$-Al$_2$O$_3$(110) exhibit even more limited diffusion and coalescence.
These findings suggest that surface dynamics is crucial to understanding nanoparticle diffusion and sintering, and support morphology engineering is a possible strategy for developing sinter-resistant supported metal catalysts.

\section{Results and Discussion}
\subsection{Active Learning Workflow}
\begin{figure}
    \includegraphics[width=1.00\textwidth]{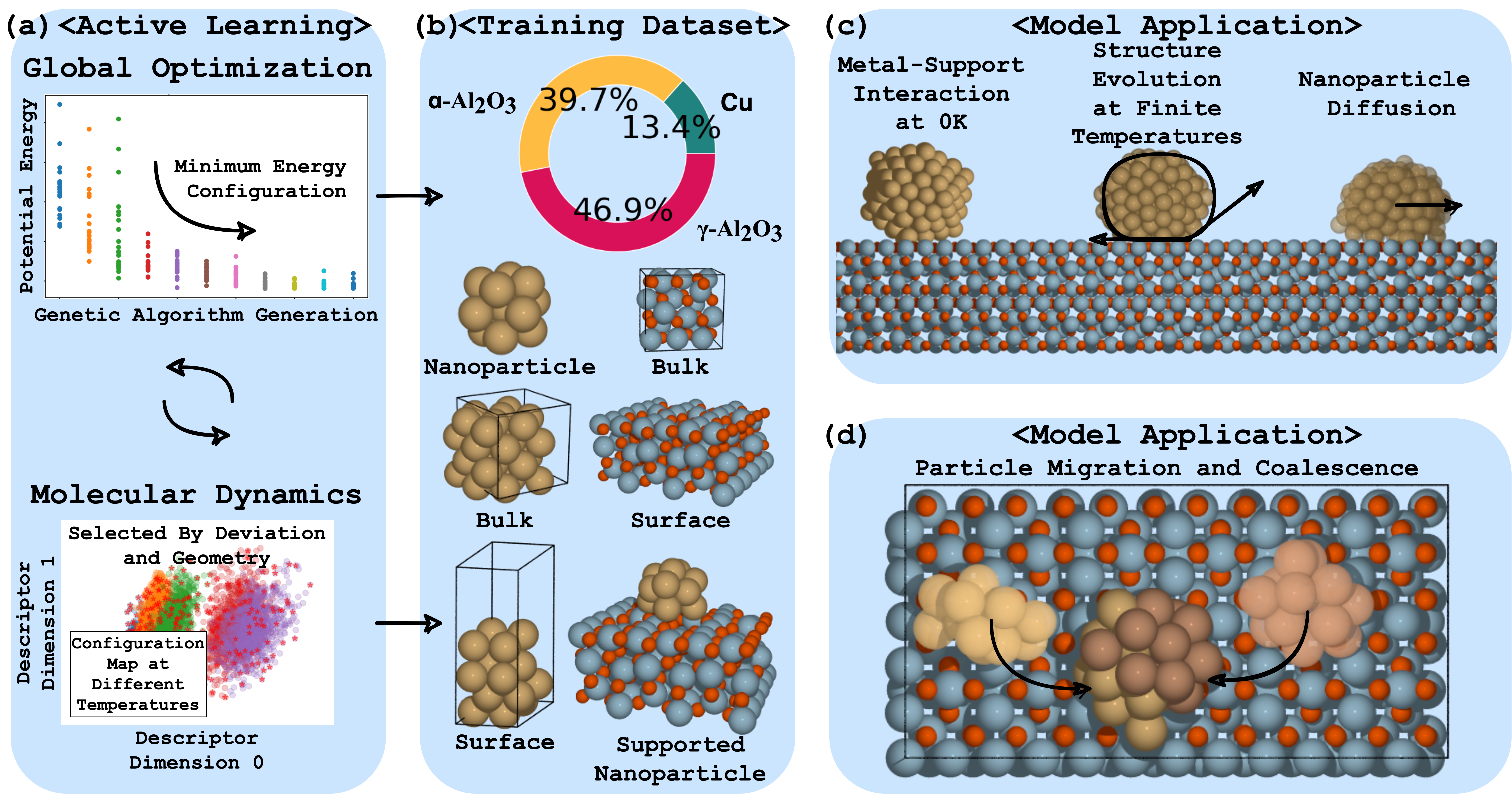}
    \caption{
        Schematic illustration of the active learning workflow. 
        (a) In each active learning cycle, two exploration methods are utilized to curate a comprehensive and compact dataset. The GA focuses on minimum energy configurations of supported nanoparticles. The MD samples structures at finite temperature, which are further sifted based on model deviation and geometry diversity. 
        (b) The training dataset covers Cu-only structures (bulk, surface, and nanoparticle), Al$_2$O$_3$ surfaces, and supported nanoparticles, including 147,464 structures in total. Our UniAlCu model has a root mean squared error of 0.004 eV/Atom in the energy and of 0.057 eV/\r{A} in the force predictions.
        Taking supported nanoparticles on $\gamma$-Al$_2$O$_3$(100) as an example, we illustrate different applications of our UniAlCu model: (c) exploring supported nanoparticle structures at zero and finite temperatures; (d) revealing atomic details of nanoparticle sintering by MD simulations over large time and length scales.
        Color code: Cu: yellow, Al: cyan and O: red. (Brown is used to distinguish the second Cu nanoparticle participating in coalescence.)
    }
    \label{fig:active_learning}
\end{figure}

To investigate the atomic-level structure evolution of Al$_2$O$_3$-supported Cu nanoparticles, we established an active learning workflow to train our DP models, namely, GamAlCu for $\gamma$-Al$_2$O$_3$, and AlpAlCu for $\alpha$-Al$_2$O$_3$.
Combining these two individual models, we created a unified DP model (UniAlCu) that is uniformly accurate for supported Cu nanoparticles on two Al$_2$O$_3$ phases (\textbf{Figure~S1} and \textbf{Section~S1}).
The active learning workflow incorporates several exploration methods as implemented in the GDPy\cite{Xu2024GDPy} package, the details of which can be found in \textbf{Section~S2}.
\textbf{Figure~\ref{fig:active_learning}a} shows the two exploration methods used through our active learning cycles. 
The GO in the flavor of genetic algorithm (GA) biases the search toward low-energy structures of supported Cu nanoparticles, providing other exploration methods with several good starting configurations.
 MD simulations at different temperatures sample a sheer volume of structures ($\gamma$- and $\alpha$-Al$_2$O$_3$ surfaces and supported Cu nanoparticles), which are further sifted on the basis of the model deviation and geometry diversity to maintain a compact data set. 
Given our focus on Al$_2$O$_3$-supported Cu nanoparticles, we organized the data set into three parts shown in \textbf{Figure~\ref{fig:active_learning}b}.
The Cu-only structures, including bulk, surface and nanoparticle systems, are taken from a previous study\cite{Wang2022JCP} while Cu$_n$/$\gamma$-Al$_2$O$_3$ and Cu$_n$/$\alpha$-Al$_2$O$_3$ ($n=$ number of atoms in the nanoparticle) structures are actively accumulated by our active learning workflow.
The $\gamma$-Al$_2$O$_3$ bulk is modeled by the structure proposed by Digne.\cite{Digne2004JCatal}
Applications outlined in \textbf{Figure~\ref{fig:active_learning}c} indicate that our UniAlCu model is capable of simulating both the global minimum configurations of supported Cu nanoparticles and their dynamic behavior at finite temperature.
Furthermore, nanoparticle sintering can be observed through long-time and large-scale simulations (\textbf{Figure~\ref{fig:active_learning}d}).
These applications will be demonstrated in the following sections.

\subsection{Minimum Energy Configurations}
The configurations of metallic nanoparticles on a supporting surface depend on the balance of MSI and intraparticle binding.
On Al$_2$O$_3$ surfaces, Cu nanoparticles can form bonds with Al and O atoms and can electrostatically interact with the substrate due to charge transfer.\cite{Pacchioni2018ChemSocRev}
To measure the interaction strength, we calculated the binding energy ($E_b$) according to \textbf{Equation~\ref{eq:binding_energy}}.
$E_b$ has two components, the adhesion energy of a nanoparticle on the surface and the energy cost of restructuring the gas phase nanoparticle to conform to the surface.
$E_b$ is a crucial descriptor that influences catalytic activity and stability, directly reflecting the MSI.\cite{Tosoni2019ChemCatChem}
Similar to $E_b$, we also evaluated the cohesive energy ($E_c$) as defined in \textbf{Equation~\ref{eq:cohesive_energy}}.
Since $E_c$ uses the atomic energy of the bulk FCC Cu as a reference, it measures the stability of a nanoparticle, supported or in the gas phase, relative to the bulk.
Thus, all nanoparticles have $E_c$ greater than zero and $E_c=0$ corresponds to the bulk limit.  

\begin{figure}[hbtp]
    \includegraphics[width=1.00\textwidth]{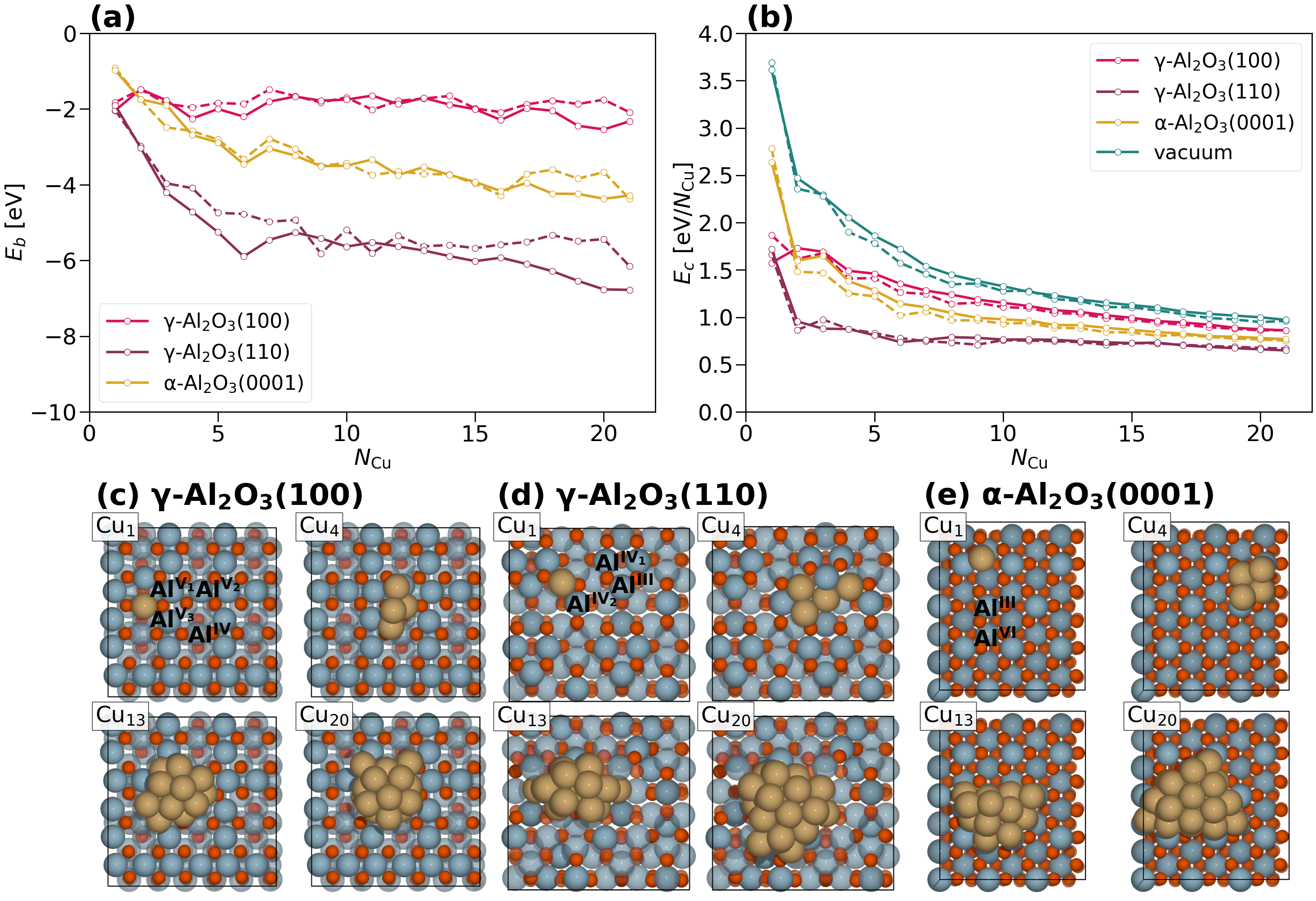}
    \caption{
        The global minimum configurations of Cu$_{1\text{-}21}$ on three Al$_2$O$_3$ surfaces are DFT-optimized structures obtained through a DP-accelerated GA search.
        The binding energy (a) and the cohesive energy (b) are compared between DP and DFT (solid and dashed lines, respectively).
        The structural evolution of supported Cu$_1$, Cu$_4$, Cu$_{13}$, and Cu$_{20}$ nanoparticles (c-e) reveals distinct growth patterns.
        The Al atoms are denoted by the superscript as their coordination numbers with O atoms. For $\gamma$-Al$_2$O$_3$(100), 
        there are three types of Al$^{V}$ on the surface and one type of Al$^{IV}$ in the subsurface. For $\gamma$-Al$_2$O$_3$(110), 
        there are two types of Al$^{IV}$ and one type of Al$^{III}$ on the surface. For $\alpha$-Al$_2$O$_3$(0001), there is one type of Al$^{III}$ on the surface and one type of Al$^{VI}$ in the subsurface.
        Nanoparticles on $\gamma$-Al$_2$O$_3$(100) are more compact than those on $\gamma$-Al$_2$O$_3$(110) and $\alpha$-Al$_2$O$_3$(0001), which is consistent with the MSI strength indicated by the binding energy.
    }
    \label{fig:binding_energy}
\end{figure}

To validate our UniAlCu model, we benchmarked $E_b$ and $E_c$ against DFT calculations on minimum energy configurations.
The detailed values of $E_b$ and $E_c$ are listed in \textbf{Table~S1-S3}.
We considered nanoparticles with a number of atoms, $N_{\mathrm{Cu}}$, ranging from 1 to 21, and used periodic supercells to model three supporting surfaces, namely $p(3\times2)$ for $\gamma$-Al$_2$O$_3$(100), $p(2\times2)$ for $\gamma$-Al$_2$O$_3$(110), and $p(3\times2\sqrt{3})$ for $\alpha$-Al$_2$O$_3$(0001).
The corresponding structures are displayed in \textbf{Figure~S2}, which also reports the O coordination number of some Al atoms as a superscripted Latin numeral. 
Each Al$_2$O$_3$ system is described by a slab whose thickness is fixed by monitoring the convergence of the calculated $E_b$, as shown in \textbf{Figure~S3}.
Fully relaxed minimum energy configurations of supported nanoparticles are found with an active learning GA protocol that efficiently explores the potential energy landscape~\cite{Xu2022ACSCatal,Han2023ACSCatal1,Han2023ACSCatal2}, as illustrated in \textbf{Figure~\ref{fig:active_learning}}. 
\textbf{Figure~\ref{fig:binding_energy}a} compares the binding energies of the nanoparticles on the three Al$_2$O$_3$ surfaces.
The negative values of $E_b$ indicate that the supported nanoparticles are always more stable than their gas-phase counterparts. 
On $\gamma$-Al$_2$O$_3$(100), $E_b$ depends weakly on $N_{\mathrm{Cu}}$, suggesting a weak size dependence of the MSI.
In contrast, on $\gamma$-Al$_2$O$_3$(110), $E_b$ takes more negative values with increasing nanoparticle size, signifying stronger size-sensitive adhesion.
The stark difference in adhesion between $\gamma$-Al$_2$O$_3$(100) and $\gamma$-Al$_2$O$_3$(110) is attributed to the local coordination environments of the Al sites on these surfaces.
The superior anchoring on $\gamma$-Al$_2$O$_3$(110) results from a unique surface structure that exposes Al$^{III}$ sites with good affinity for Cu atoms, which, in combination with geometric matching, strengthen the adhesion of Cu nanoparticles.
Similar to $\gamma$-Al$_2$O$_3$(110), $E_b$ on $\alpha$-Al$_2$O$_3$(0001) decreases with nanoparticle size with values that are intermediate between $\gamma$-Al$_2$O$_3$(100) and $\gamma$-Al$_2$O$_3$(110), where our computed $E_b$ for a single Cu adatom agrees with a previously reported value.\cite{Hernandez2005SurfSci}
Undercoordinated Al sites, particularly in penta-coordinated Al$^V$ ($\gamma$-Al$_2$O$_3$(100)) and planar Al$^{III}$ ($\gamma$-Al$_2$O$_3$(110) and $\alpha$-Al$_2$O$_3$(0001)) environments, play a crucial role in the initial anchoring and subsequent growth of Cu nanoparticles.
Our UniAlCu model accurately captures the $E_b$ trends across different surfaces and nanoparticle sizes, closely matching the DFT results.
Despite minor quantitative deviations for some nanoparticle sizes, the model reliably reproduces the relative binding strengths, demonstrating its predictive power in ranking the MSI strength.
By further analysis of the surface structures, the displacement of Al atoms on the surface is found to reflect the binding strength between the nanoparticle and the surface (\textbf{Figure~S4}).
A large displacement of surface Al atoms in the z-direction is observed on $\gamma$-Al$_2$O$_3$(110) and $\alpha$-Al$_2$O$_3$(0001) while minimal displacement occurs on $\gamma$-Al$_2$O$_3$(100).
In particular, this displacement increases with the nanoparticle size on $\alpha$-Al$_2$O$_3$(0001) as more Al$^{III}$ atoms are stabilized by the interaction with neighboring Cu atoms, leading to a more negative $E_b$.

The $E_c$ of Cu nanoparticles on the three Al$_2$O$_3$ surfaces are shown in \textbf{Figure~\ref{fig:binding_energy}b}.
Smaller nanoparticles generally exhibit larger $E_c$, implying a propensity to sinter into larger nanoparticles, leading to catalyst deactivation over time.\cite{Monzon2003ApplCatalA}
On $\gamma$-Al$_2$O$_3$(100), $E_c$ smoothly approaches the bulk limit with increasing size. Conversely, on $\gamma$-Al$_2$O$_3$(110), a sharp drop in $E_c$ from Cu$_1$ to Cu$_2$ is followed by a plateau, suggesting similar stability for a range of intermediate sizes. 
Likewise, we observed a sharp drop in $E_c$ from Cu$_1$ to Cu$_2$, followed by a smooth transition on $\alpha$-Al$_2$O$_3$(0001).
Notably, on all three Al$_2$O$_3$ surfaces, $E_c$ of Cu nanoparticles is significantly lower than that of the gas phase nanoparticles, especially for small sizes. 
This highlights the crucial role of MSI in stabilizing dispersed nanoparticles that would be unstable in the gas phase.
As size increases, the difference between supported and isolated nanoparticles decreases, consistent with the trend of $E_b$.
Our UniAlCu model shows excellent agreement with DFT in predicting both absolute cohesive energies and size-dependent trends, further validating its accuracy.

The configurations of Cu$_1$, Cu$_4$, Cu$_{13}$ and Cu$_{20}$ nanoparticles on the three Al$_2$O$_3$ surfaces (\textbf{Figure~\ref{fig:binding_energy}c-e}) reveal distinct growth patterns. 
On $\gamma$-Al$_2$O$_3$(100), the nanoparticles transition from 2D planar to 3D compact geometries as they grow, driven by the increasing Cu-Cu interaction relative to the Cu-surface interaction (\textbf{Figure~S5}).
This shift rationalizes the gradual weakening of $E_b$ and $E_c$ with size.
In contrast, on $\gamma$-Al$_2$O$_3$(110), the nanoparticles grow around an Al$^{III}$ site up to Cu$_{10}$, following an epitaxial growth along the x-axis to bind to another Al$^{III}$ site up to Cu$_{17}$, and then more Cu atoms are added along the y axis (\textbf{Figure~S6}).
From Cu$_2$ to Cu$_{11}$ the nanoparticles include one Cu atom in the subsurface below Al$^{III}$, while from Cu$_{12}$ to Cu${_{21}}$ there are two such atoms, similar to the behavior of supported Pt nanoparticles reported in a recent DFT study.\cite{Khan2023JPCC}
On $\alpha$-Al$_2$O$_3$(0001), the nanoparticles tend to stabilize around Al$^{III}$ atoms by pulling them out of the surface plane and converting from planar to the trigonal pyramidal geometry of the surface AlO$_3$ units beneath the nanoparticle (\textbf{Figure~S7}). 
In these configurations, one or more Al$^{III}$ atoms can coordinate with several Cu atoms, consistent with the formation of metallic Cu-Al bonds. In the special case of Cu$_1$, this does not occur as the Cu-Al interaction is too weak, and the minimum energy configuration maximizes the interaction of Cu with the O atoms above a subsurface Al$^{VI}$ site. 
The shape of the supported nanoparticles correlates with their radius of gyration ($R_g$) shown in \textbf{Figure~S8}, which increases steadily with the size of the nanoparticle, indicating an increasing spread of the Cu atoms in the surface plane.
For all sizes, the nanoparticles on $\gamma$-Al$_2$O$_3$(100) have a smaller $R_g$ compared to $\gamma$-Al$_2$O$_3$(110) and $\alpha$-Al$_2$O$_3$(0001), indicating a more compact shape consistent with a weaker MSI. 

\subsection{Nanoparticles at Finite Temperature}
\begin{figure}
    \includegraphics[width=1.00\textwidth]{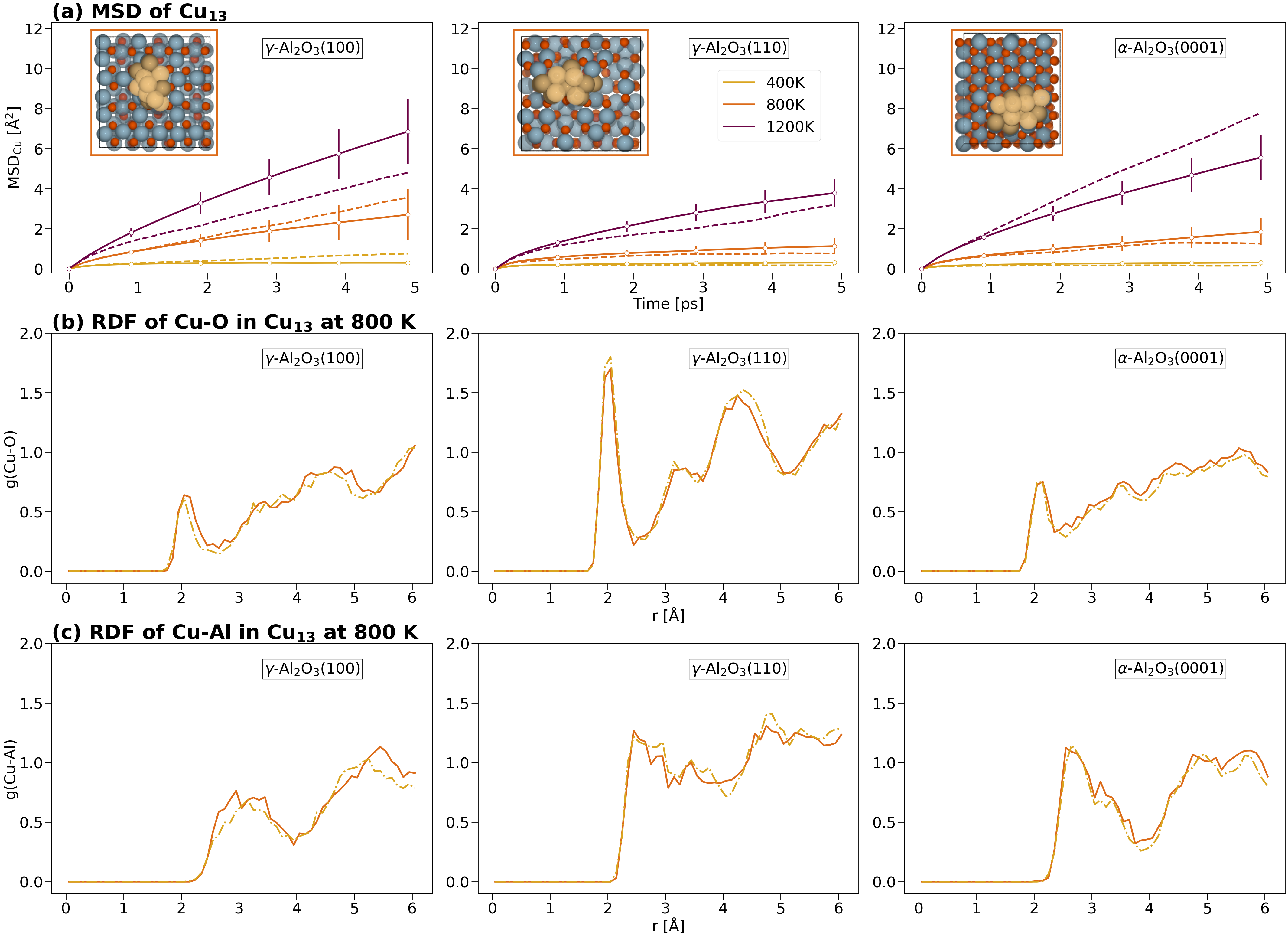}
    \caption{
        (a) The MSD$_\mathrm{Cu}$ of Cu$_{13}$ on $\gamma$-Al$_2$O$_3$(100), $\gamma$-Al$_2$O$_3$(110) and $\alpha$-Al$_2$O$_3$(0001) is evaluated by NVT simulations at 400 K, 800 K, 1200 K.
        The solid and dashed lines are DP and DFT results, respectively.
        The vertical bars indicate the standard errors from 20 independent DPMD trajectories.
        The insets are snapshots at the elapsed time of 20 ps in the AIMD simulations at 800 K.
        The light-yellow colored Cu atoms are those on top of the nanoparticle at 0 ps.
        It is clear that the nanoparticles $\gamma$-Al$_2$O$_3$(100) and $\alpha$-Al$_2$O$_3$(0001) undergo significant structual relaxation.
        (b)-(c) The RDFs of Cu-O and Cu-Al pairs in Cu$_{13}$ on three Al$_2$O$_3$ surfaces at 800 K.
        The solid and dashed lines are DP and DFT results, respectively.
        The stronger Cu-O peak for $\gamma$-Al$_2$O$_3$(110) suggests the formation of stronger Cu-O bonds.
        The Cu-Al peak for $\gamma$-Al$_2$O$_3$(100) and $\alpha$-Al$_2$O$_3$(0001) indicates the formation of metallic bonds between Cu and Al atoms.
    }\label{fig:single_dynamics}
\end{figure}

To benchmark the performance of our UniAlCu model at finite temperature, we compare the mean squared displacement (MSD) of the Cu atoms (MSD$_\mathrm{Cu}$ defined in \textbf{Equation~\ref{eq:msd}}) extracted from DPMD and AIMD simulations of Cu$_{13}$ on $p(3\times2)$ $\gamma$-Al$_2$O$_3$(100), $p(2\times2)$ $\gamma$-Al$_2$O$_3$(110), and $p(3\times2\sqrt3)$ $\alpha$-Al$_2$O$_3$(0001) surfaces at 400 K, 800 K, and 1200 K (\textbf{Figure~\ref{fig:single_dynamics}a}).
The trajectories start from the global minimum configurations of nanoparticle plus substrate with initial atomic velocities drawn from the Maxwell-Boltzmann distributions associated with the different atomic species at the simulation temperature.
The trajectories run for 20 ps in the NVT ensemble with temperature controlled by a Langevin thermostat to ensure proper equilibration.\cite{Korpelin2022JPCL}
For each temperature and surface orientation, we ran one AIMD trajectory and 20 independent DPMD trajectories.
MSD$_\mathrm{Cu}$ in a time interval of 5 ps is extracted from the trajectories by window averaging (see details in \textbf{Section~S4.1}) and reported in \textbf{Figure~\ref{fig:single_dynamics}a}.
In the DPMD case, the standard errors of the simulated MSD$_\mathrm{Cu}$ are estimated from the 20 independent trajectories and shown in the same figure as the vertical bars.
Our UniAlCu model reproduces AIMD within the error bars of DPMD.
At 400 K, diffusion is negligible, and displacements reflect vibrational motion.
At higher temperatures (800 K and 1200 K), diffusion is activated and larger MSD$_\mathrm{Cu}$ values, which grow approximately linearly with time, are observed on the three Al$_2$O$_3$ surfaces.
This behavior is attributed to internal diffusion within the nanoparticles because no significant displacements of the center of mass (COM) are observed on the short timescale of these simulations. 
In agreement with the indication of $E_b$, the strong MSI on $\gamma$-Al$_2$O$_3$(110) hinders Cu mobility.
However, at high temperature, Cu$_{13}$ on $\alpha$-Al$_2$O$_3$(0001) shows comparable mobility to that observed on $\gamma$-Al$_2$O$_3$(100) in spite of the significantly larger $E_b$ on the $\alpha$-surface than on the $\gamma$-surface (-3.71 eV vs -1.72 eV at the DFT level).
A similar behavior is observed in the simulations of the supported Cu$_4$ and Cu$_{20}$ nanoparticles (\textbf{Figure~S10} and \textbf{Figure~S12}). 
The structures in the insets of \textbf{Figure~\ref{fig:single_dynamics}a} are snapshots at the elapsed time of 20 ps in the AIMD simulations at 800 K.
They reveal that even in the absence of COM diffusion, the nanoparticles on $\gamma$-Al$_2$O$_3$(100) and $\alpha$-Al$_2$O$_3$(0001) undergo significant structural relaxation by exploring low-energy configurations distinct from the ground-state configuration.
\textbf{Figure~\ref{fig:single_dynamics}b} and \textbf{Figure~\ref{fig:single_dynamics}c} compare the Cu-O and Cu-Al radial distribution functions (RDFs) extracted from the AIMD and DPMD simulations of Cu$_{13}$ on the three surfaces at 800 K.
The first peak of the Cu-O and Cu-Al RDFs provides information on the bonding of the nanoparticle with the substrate.
The Cu-O peak is stronger and sharper on $\gamma$-Al$_2$O$_3$(110), suggesting the formation of stronger Cu-O bonds on that surface.
The Cu-Al peak is well defined on $\gamma$-Al$_2$O$_3$(110) and on $\alpha$-Al$_2$O$_3$(0001), where it is associated with metallic bonds between Cu and Al$^{III}$ atoms that are pulled out of the $\alpha$-Al$_2$O$_3$(0001) surface plane.
Comparisons for Cu$_1$, Cu$_4$, Cu$_{13}$, and Cu$_{20}$ (\textbf{Figure~S13-S16}) further support the good agreement between AIMD and DPMD based on our UniAlCu model. 

Moreover, we validated our UniAlCu model by estimating the melting temperature $T_m$ of the bulk Cu and gas phase nanoparticles.
The bulk $T_m$ was obtained by simulating the coexistence of the liquid and solid phases, finding a value of 1207 K for $T_m$ (\textbf{Figure~S17}), close to an earlier \textit{ab initio} prediction\cite{Vovcadlo2004JCP}and in fair agreement with the experimental value of 1358 K.
We estimated the size dependence of the melting temperatures of Cu nanoparticles in vacuo from the Lindeman index, as shown in \textbf{Figure~S18}, finding a $T_m \propto N_\mathrm{Cu}^{-1/3}$ dependence on the number of Cu atoms in the nanoparticle, in good agreement with a recent study.\cite{Weinreich2020JPCC}
In addition, we used configuration maps based on Smooth Overlap of Atomic Positions (SOAP) vectors\cite{Bartok2013PRB,De2016PCCP} to rank the configurations visited in short AIMD trajectories and those visited in much longer DPMD trajectories that exhibit COM diffusion, finding substantial overlap of the respective maps (\textbf{Figure~S19}).
This shows that long DPMD simulations sample configurations similar to those encountered in short AIMD simulations but with displaced COM positions.
We conclude that our UniAlCu model is capable of accurately reproducing geometrical, local bonding, and dynamical relaxation trends, suggesting that the model should also be predictive of the behavior of larger supported nanoparticles on longer time scales.  

\begin{figure}[hbtp]
    \includegraphics[width=1.00\textwidth]{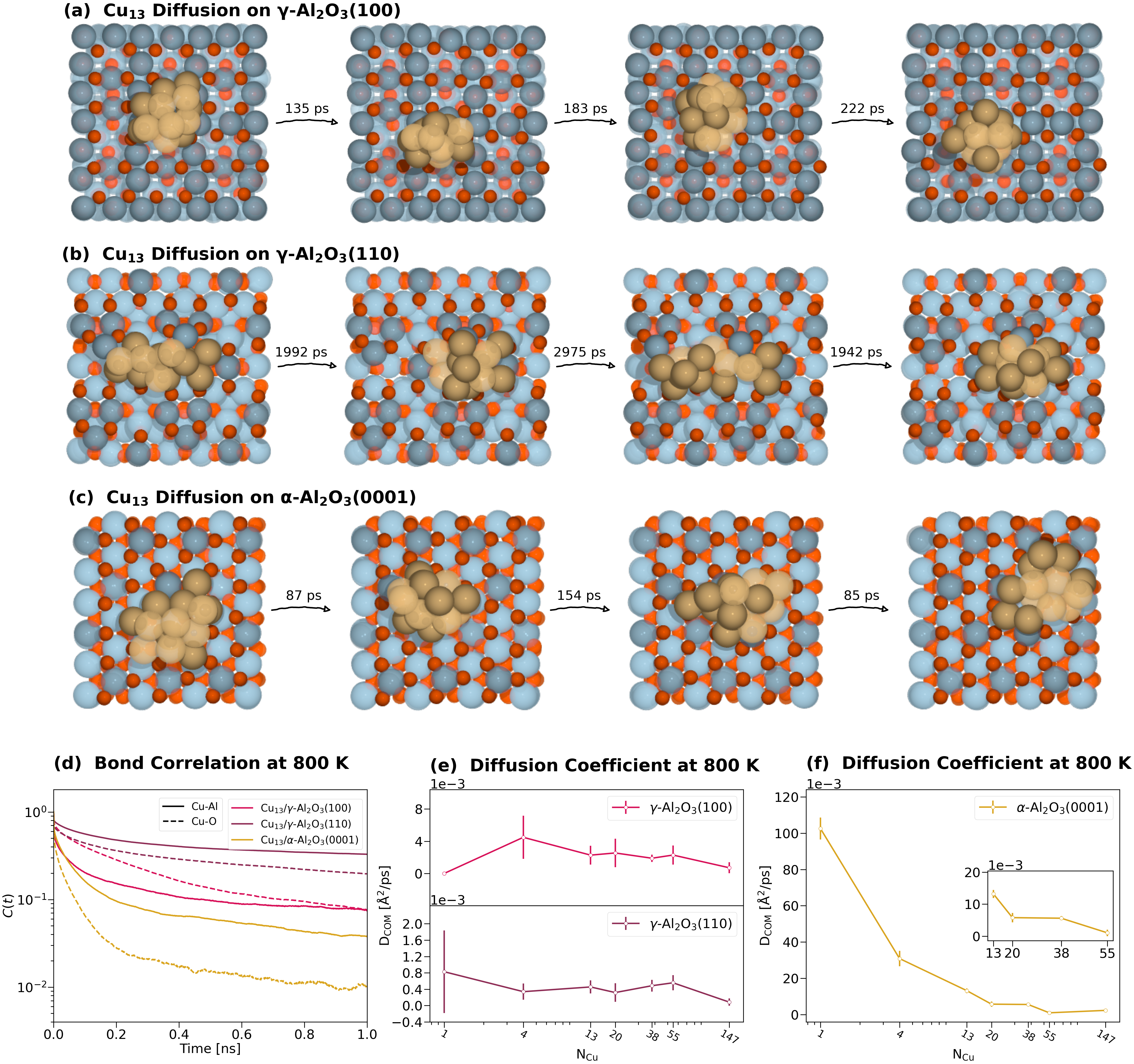}
    \caption{
        (a)-(c) The snapshots of three successive configurations with displaced COMs of Cu$_{13}$ on three Al$_2$O$_3$ surfaces.
        The light-yellow colored Cu atoms are those on the top of the nanoparticle in the first snapshot.
        The surface Al atoms are colored based on their heights (higher Al atoms are darker) and the subsurface Al and O atoms are painted pale.
        The magnitude of the elapsed time between the snapshots is quite different on three surfaces, about 100 ps on $\gamma$-Al$_2$O$_3$(100) and $\alpha$-Al$_2$O$_3$(0001) while about 2.0 ns on $\gamma$-Al$_2$O$_3$(110).
        (d) The diffusion is controlled mainly by Cu-O interactions as the Cu-O correlations decay more slowly than the Cu-Al correlations.
        The rapid decay in both Cu-Al and Cu-O correlations on $\alpha$-Al$_2$O$_3$(0001) elucidates the fast diffusion.
        (e)-(f) The diffusion coefficients of different nanoparticle sizes on $\gamma$- and $\alpha$-Al$_2$O$_3$ surfaces.
        The vertical bars indicate the standard errors from 5 independent DPMD trajectories.
        The diffusion coefficient on $\alpha$-Al$_2$O$_3$(0001) decreases significantly with nanoparticle size while it is still larger than those on $\gamma$-Al$_2$O$_3$(100) and $\gamma$-Al$_2$O$_3$(110).
    }
    \label{fig:diffusion}
\end{figure}

To understand the long-time diffusion behavior of supported nanoparticles, we performed 10-ns-long DPMD simulations for a range of nanoparticle sizes from Cu$_1$ up to Cu$_{675}$.
The adopted structure models are shown in \textbf{Figure~S20-S22} and the details of the simulation protocol is given in \textbf{Section~7.1}.
Long DPMD simulations show that the diffusion dynamics of the COM of supported nanoparticles is slow and may involve the separation of time scales between the internal dynamics of a nanoparticle and its diffusional motion.
This behavior is illustrated for a Cu$_{13}$ nanoparticle in \textbf{Figures~\ref{fig:diffusion}a,b,c}, which show snapshots of three successive displaced configurations along DPMD trajectories.
The time evolution of the absolute value of the COM displacement is reported in \textbf{Figure~S23}, \textbf{Figure~S24}, \textbf{Figure~S25} for $\gamma$-Al$_2$O$_3$(100), $\gamma$-Al$_2$O$_3$(110), and $\alpha$-Al$_2$O$_3$(0001), respectively.
It shows that on the two $\gamma$ surfaces the COM dynamics combines fluctuations that do not contribute to diffusion with relatively large occasional jumps of about 2.5 \r{A} on $\gamma$-Al$_2$O$_3$(100) and of about 2.0 \r{A} on $\gamma$-Al$_2$O$_3$(110).
The magnitudes of the jumps match approximately the distance between nearest-neighbor Al atoms on the two surfaces.
The time elapsed between the snapshots of \textbf{Figures~\ref{fig:diffusion}a,b} is quite different on the two surfaces, greater than 100 ps on $\gamma$-Al$_2$O$_3$(100), and greater than 2.0 ns on $\gamma$-Al$_2$O$_3$(110), where diffusion is hampered by strong bonds between Cu and Al$^{III}$ and Cu and O atoms.
Diffusion on $\alpha$-Al$_2$O$_3$(0001) is facilitated by dynamic MSI and is faster than on $\gamma$-Al$_2$O$_3$(100).
Interestingly, the separation of time scales observed on the two $\gamma$ surfaces is not evident on the $\alpha$ surface, as \textbf{Figure~S25} depicts an evolution that proceeds by a succession of small displacements on the time scale of the COM fluctuations, promoted by the motion in and out of the surface plane of the Al$^{III}$ atoms adjacent to the nanoparticle.
The above analysis is further corroborated by the Cu-Al and Cu-O bond correlation functions ($C(t)$, \textbf{Equation~\ref{eq:bond_correlation}}) reported for Cu$_{13}$ on the three surfaces (\textbf{Figure~\ref{fig:diffusion}d}).
In defining $C(t)$, only the Al and O atoms in the top surface layer, which interact directly with the nanoparticle, were considered.
In addition to O, these include Al$^{V_{1}}$, Al$^{V_{2}}$, Al$^{V_{3}}$, and Al$^{IV}$ on $\gamma$-Al$_2$O$_3$(100); Al$^{IV_{1}}$, Al$^{IV_{2}}$ and Al$^{III}$ on $\gamma$-Al$_2$O$_3$(110); Al$^{III}$ and Al$^{VI}$ on $\alpha$-Al$_2$O$_3$(0001)).
Unlike $E_b$, which is calculated from the relaxed structures at 0 K, the bond correlation reflects MSI in a dynamic context.
The Cu-Al and Cu-O bond correlation functions decay slowly on $\gamma$-Al$_2$O$_3$(110), as expected from strong binding and slow diffusion.
On $\gamma$-Al$_2$O$_3$(100), diffusion is faster and the bond correlation functions decay more rapidly.
Interestingly, the Cu-O correlations decay more slowly than the Cu-Al correlations, suggesting that the diffusion is controlled mainly by Cu-O interactions. 
By contrast, on $\alpha$-Al$_2$O$_3$(0001), the Cu-O bond correlation function decays faster than on $\gamma$-Al$_2$O$_3$(100) as the dynamic Al$^{III}$ layer that facilitates diffusion and hinders close contacts between Cu and O atoms. 

The COM diffusion coefficients ($D_\mathrm{COM}$) defined in \textbf{Equation~\ref{eq:diffusion_coefficient}} are extracted from five 10-ns-long DPMD simulations for a range of nanoparctile sizes up to Cu$_{147}$ (\textbf{Figures~\ref{fig:diffusion}e,f}).
The standard errors of the simulated $D_\mathrm{COM}$ are estimated from the five independent DPMD trajectories and are shown in the same figure as the vertical bars.
The $D_\mathrm{COM}$ values for Cu$_{309}$ and Cu$_{675}$ are listed in \textbf{Table~S4}.
On the two $\gamma$-Al$_2$O$_3$ surfaces diffusion is almost independent of the nanoparticle size, with $D_\mathrm{COM}$ smaller on $\gamma$-Al$_2$O$_3$(110) than on $\gamma$-Al$_2$O$_3$(100), consistent with the indication of $E_b$.
On $\alpha$-Al$_2$O$_3$(0001), the diffusion is significantly faster for a single adatom and small nanoparticles, but decreases rapidly with size, while always remaining somewhat faster than on $\gamma$-Al$_2$O$_3$(100) despite the larger binding of the nanoparticles in $\alpha$-Al$_2$O$_3$(0001) than in $\gamma$-Al$_2$O$_3$(100), consistent with the important role that dynamic MSI plays on the $\alpha$-Al$_2$O$_3$(0001). 
In the case of Cu$_1$, we performed Climbing Image Nudged Elastic Band (CI-NEB)\cite{Henkelman2000JCP} calculations at zero temperature to estimate the diffusion barriers among different surface sites.
The diffusion barrier of Cu$_1$ on $\alpha$-Al$_2$O$_3$(0001) is of only about 0.15 eV (\textbf{Figure~S26}), much smaller than the barriers on $\gamma$-Al$_2$O$_3$(100) and $\gamma$-Al$_2$O$_3$(110) (\textbf{Figure~S27} and \textbf{Figure~S28}).

\subsection{Nanoparticle Sintering Dynamics}
\begin{figure}
    \includegraphics[width=1.00\textwidth]{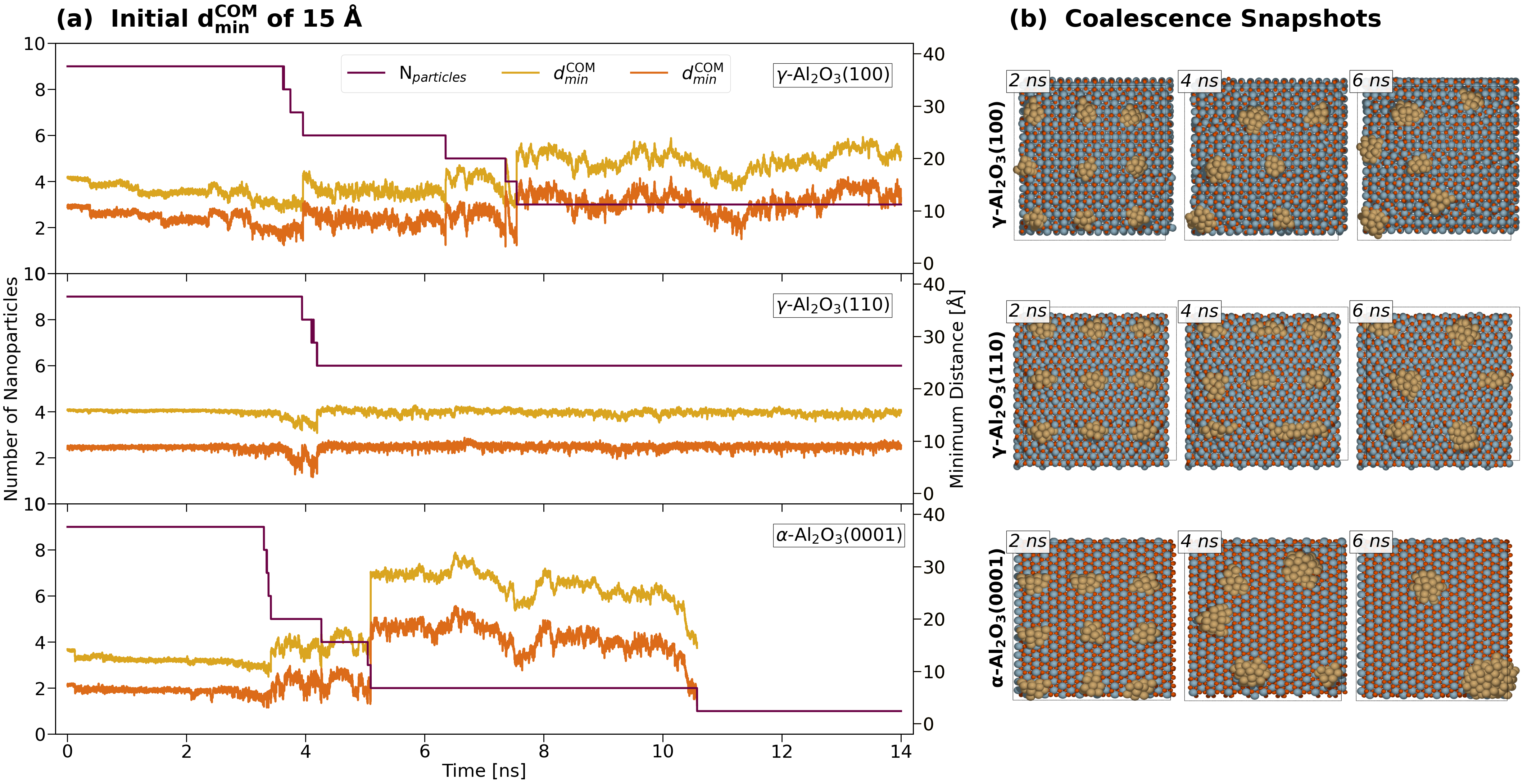}
    \caption{
        (a) Nine Cu$_{13}$ nanoparticles are placed on the surface with an initial inter-nanoparticle distance of about 15 \r{A}. During the first 2 ns at 400 K, negligible nanoparticle diffusion on all three surfaces is observed. Coalescence starts to occur as the temperature is gradually increased to 800 K. All nine Cu$_{13}$ nanoparticles merge into a single Cu$_{117}$ nanoparticle on $\alpha$-Al$_2$O$_3$(0001), while the largest nanoparticle found after 14 ns of simulation is Cu$_{39}$ and Cu$_{26}$ on $\gamma$-Al$_2$O$_3$(100) and $\gamma$-Al$_2$O$_3$(110), respectively. 
        (b) The snapshots exhibit that the nanoparticle diffusion is pronounced on $\gamma$-Al$_2$O$_3$(100) and $\alpha$-Al$_2$O$_3$(0001) while it is significantly limited on the $\gamma$-Al$_2$O$_3$(110) surface and thus coalescence only occurs between row-aligned nanoparticles by the epitaxial reconfiguration.
        The overall sintering becomes more difficult with the increasing nanoparticle size since the nanoparticle mobility is reduced.
    }
    \label{fig:sintering}
\end{figure}

After gaining insights into the static and dynamic behavior of isolated supported nanoparticles, we delve into the intricate process of nanoparticle sintering on the three Al$_2$O$_3$ surfaces considered in the previous sections.
For that purpose, we strategically placed on each surface nine Cu$_{13}$ nanoparticles in their global minimum configuration at equal distances of approximately 15 \r{A} (\textbf{Figure~S30}).
Then, we perform DPMD simulations controlled by a Langevin thermostat according to the following schedule. First, we equilibrate the systems at 400 K for 2 ns, then heat them at a constant rate of 0.2 K/ps for 2 ns, and finally, we run equilibrium NVT trajectories for 10 ns at 800 K.
We employ the depth-first search (DFS)\cite{Even2011GraphAlgorithms} graph alogorithm to find nanoparticles in the system and monitor Cu-Cu connectivity, nanoparticle count, minimum inter-COM distance ($d_{min}^{\mathrm{COM}}$, \textbf{Equation~\ref{eq:min_com_distance}}), and minimum inter-Cu distance ($d_{min}^{\mathrm{Cu}}$, \textbf{Equation~\ref{eq:min_copper_distance}}) throughout the simulation (\textbf{Figure~\ref{fig:sintering}a}).
These metrics serve as indicators of possible coalescence.
During the initial 2 ns at 400 K, the nanoparticle count does not change on the three surfaces, while $d_{min}^{\mathrm{COM}}$ fluctuates around 15 \r{A}, as expected from the negligible diffusion at this temperature.
The initial minimum distance between Cu atoms belonging to different nanoparticles ($d_{min}^{\mathrm{Cu}}$) on the three surfaces is close to or greater than our UniAlCu model cutoff of 8 \r{A}, indicating that interparticle interactions should be negligible and motion should be primarily controlled by MSI.
As the temperature increases during the heating protocol, we observe a first coalescence event on $\alpha$-Al$_2$O$_3$(0001) at a temperature of around 600 K,  evidenced by a drop in $d_{min}^{\mathrm{COM}}$ and $d_{min}^{\mathrm{Cu}}$.
Other coalescence events follow, until a single Cu$_{117}$ nanoparticle forms after approximately 6 ns at 800 K.
A similar behavior is observed on $\gamma$-Al$_2$O$_3$(100) but here coalescence becomes rarer with increasing nanoparticle sizes, and three nanoparticles are left on this surface at the end of 14 ns of simulation.
Coalescence is even more difficult on $\gamma$-Al$_2$O$_3$(110) due to the reduced nanoparticle mobility on this surface.
Therefore, six nanoparticles are present after 6 ns, and their number does not change in the remaining 8 ns of simulation.
The snapshots in \textbf{Figure~\ref{fig:sintering}b} suggest that sintering on $\alpha$-Al$_2$O$_3$(0001) and $\gamma$-Al$_2$O$_3$(100) is driven by nanoparticle diffusion while epitaxial reconfiguration promotes coalescence between row-aligned nanoparticles on $\gamma$-Al$_2$O$_3$(110).
Increasing the initial separation between the particles to 20 \r{A} and 30 \r{A}, respectively, for $d_{min}^{\mathrm{COM}}$, that is, 15 and 20 \r{A} for $d_{min}^{\mathrm{Cu}}$, further suppresses sintering on the three surfaces, as illustrated in \textbf{Figure~S31}. For
$d_{min}^{\mathrm{COM}}=30$ \r{A}, coalescence is only observed on $\alpha$-Al$_2$O$_3$(0001).
Compared to nanoparticles on the two $\gamma$-Al$_2$O$_3$ surfaces (\textbf{Figure~S32} and \textbf{Figure~S33}), those on $\alpha$-Al$_2$O$_3$(0001) (\textbf{Figure~S34}) exhibit more pronounced distance fluctuations, underscoring the pivotal role of diffusion in the sintering process.
A previous experimental study on Pt/C catalysts demonstrated that the sintering via migration and coalescence can be suppressed greatly by increasing the initial distance between nanoparticles,\cite{Yin2021NatCommun} which is consistent with the observations from our DPMD simulations for Cu nanoparticles on Al$_2$O$_3$ surfaces.

In the presence of large particles with negligible mobility, the sintering process may occur by Ostwald ripening, a mechanism in which a nanoparticle grows by incorporating adatoms diffusing on the surface.
Cu$_1$ on $\gamma$-Al$_2$O$_3$(100) exhibits essentially zero diffusion at 800 K once it is in the global minimum structure, in which it occupies the site of a surface Al$^{V}$ atom.  
In such a case, for large enough particles, sintering should be largely inhibited, since Ostwald ripening will be suppressed, as well as migration and coalescence.
On $\gamma$-Al$_2$O$_3$(110), Cu$_1$ shows a diffusion comparable to that of other nanoparticles, suggesting that Ostwald ripening should be as effective as migration and coalescence. 
However, in general, sintering should be slow on that surface as a result of the limited diffusion.
On $\alpha$-Al$_2$O$_3$(0001), Cu$_1$ diffuses much faster than larger nanoparticles, suggesting that sintering should not be inhibited even for nanoparticles large enough so that Ostwald ripening becomes dominant.
Based on our findings, the pristine $\gamma$-Al$_2$O$_3$(110) should be the best sinter-resistant support among the three surfaces investigated.

\section{Conclusion}
In this work, we have developed an active learning workflow, incorporating GA search and MD sampling, to train a robust and transferable MLIP model for supported nanoparticles, which can be used to investigate the supported nanoparticle dynamics through simulations with extended time and length scales.
Our UniAlCu model for Cu nanoparticles supported on three Al$_2$O$_3$ surfaces ($\gamma$-Al$_2$O$_3$(100), $\gamma$-Al$_2$O$_3$(110), and $\alpha$-Al$_2$O$_3$(0001)) achieves remarkable accuracy with root mean squared errors of 0.004 eV/atom in energy and 0.057 eV/\r{A} in force. 
With careful benchmarks on the key descriptors of supported nanoparticles ($E_b$, $E_c$, MSD$_\mathrm{Cu}$, and RDF), we demonstrate that our model is capable of simulating the low-energy configurations of Cu nanoparticles on three Al$_2$O$_3$ surfaces, as well as their dynamic behavior at finite temperature, in agreement with AIMD. 

With force field-like efficiency and DFT-level accuracy, our simulations reveal striking differences in Cu nanoparticle stability and mobility patterns across the three Al$_2$O$_3$ surfaces, which would be unattainable with regular DFT-based structural minimizations and AIMD. 
According to global minimum configurations from extensive structural searches, Cu nanoparticles bind significantly more strongly on $\gamma$-Al$_2$O$_3$(110) and $\alpha$-Al$_2$O$_3$(0001) than on $\gamma$-Al$_2$O$_3$(100) surfaces due to the Cu affinity of Al$^{III}$ sites.
Furthermore, we evaluated the diffusion coefficients of nanoparticles ($D_\mathrm{COM}$) of various sizes and found that Cu nanoparticles on $\alpha$-Al$_2$O$_3$(0001) exhibit much faster diffusion than those on $\gamma$-Al$_2$O$_3$(100) at finite temperature despite stronger bindings at 0 K.
The contrast between fast diffusion and strong binding is reconciled by the dynamic MSI characterized by the bond correlation functions, where the Cu-Al and Cu-O bonds in Cu nanoparticles on $\alpha$-Al$_2$O$_3$(0001) relax remarkably fast. 
Extended MD simulations of nine Cu$_{13}$ nanoparticles on different Al$_2$O$_3$ surfaces reveal distinct sintering behaviors, with complete coalescence on $\alpha$-Al$_2$O$_3$(0001), moderate coalescence on $\gamma$-Al$_2$O$_3$(100), and limited nanoparticle diffusion and coalescence on $\gamma$-Al$_2$O$_3$(110) within 14 ns at 800 K.

The insights gained from this work reinforce the importance of dynamics in understanding supported nanoparticle sintering mechanisms and stress that descriptors that include dynamic effects are indispensable for the screening of sinter-resistant catalysts.
In the future, our active learning protocol could be applied to investigate various combinations of metals and supports as well as more intricate systems, for example, supported nanoparticles under reaction conditions, paving the way for the rational design of sinter-resistant catalysts.

\section{Methods}\label{sec:methods}
\subsection{Density Functional Theory}
All DFT calculations are performed with the Vienna Ab initio Simulation Package (VASP),\cite{Kresse1996ComputMaterSci,Kresse1996PRB} using the Perdew-Burke-Ernzerhof (PBE) approximation for the exchange-correlation functional.\cite{Perdew1996PRL}
The Projector augmented wave method (PAW)-PBE \cite{Kresse1999PRB} pseudopotential with a cutoff energy of 400 eV was adopted to describe the core-valence interaction of the electrons.
To sample the Brilluoin zone we used $\Gamma$-centered uniform $k$-point meshes with a spacing of 0.04 \r{A}$^{-1}$ in each direction of the reciprocal lattice.\cite{Monkhorst1976PRB}

\subsection{Machine Learning Interatomic Potential}
We trained an MLIP based on the DP architecture\cite{Zhang2018PRL,Zhang2018NeurIPS} using the DeePMD-kit package.\cite{Wang2018CPC,Zeng2023JCP} The DP scheme employs neural networks to map the local atomic environment onto the atomic energy. The local atomic environment is defined by neighboring atoms within a cut-off radius of 8 \r{A}. The `se\_e2\_a' descriptor is adopted to represent the atomic local environment with an embedding net of $25\times50\times100$. The fitting net from the descriptor to the atomic energy is $240\times240\times240$. The details of the training and testing data sets can be found in \textbf{Section~S1}.

\subsection{Global Optimization}
To sample the low-energy configurations of the supported nanoparticles, we used the GA search\cite{Vilhelmsen2012PRL,Vilhelmsen2014JCP} implemented in ASE\cite{Larsen2017JPhysCondensMatter} and interfaced with the active learning loop in GDPy\cite{Xu2024GDPy}, which has been successfully applied to study several catalyst surface systems in previous work.\cite{Lee2022JPCC,Xu2022ACSCatal,Han2023ACSCatal1,Han2023ACSCatal2}
The search is organized into generations. The structures in the initial generation are randomly generated while the ones in the following generations are constructed by crossover and mutation.
The crossover operation generates a new structure by a combination of two parent structures by cut-and-splice.\cite{Deaven1995PRL}
The mutation operation is performed on the offspring structures by applying with equal probability either a mirror reflection or a rattle position displacement.
Once a structure is generated, it is allowed to relax and its fitness is estimated from its total energy.
Since the GA protocol may generate several duplicate structures, we developed a coordination-number-dependent comparison method to distinguish low-energy structures, the details of which can be found in \textbf{Section~S2.1}.
Structure sampling by GA is carried out in an active learning way. Within each cycle, the scheme searches for 10 generations with a population size of 20, thus generating 200 structures, which are added to the dataset.
To avoid excluding from the search good guesses for the global minimum structures, we further enlarged the search with 20 generations and a population size of 50. Then,
the global minimum was found within the 50 DFT-minimized low-energy structures selected from all the DP-sampled minima.

From the global minimum configurations of the supported nanoparticles, we estimate the stability of the nanoparticles with $n$ atoms at 0 K with two descriptors, the binding energy $E_b$ and the cohesive energy $E_c$, which are defined as
\begin{equation}\label{eq:binding_energy}
    E_b = E_{{\mathrm{Cu}_n}/surf} - E_{surf} - E_{{\mathrm{Cu}_n}}
\end{equation}
and
\begin{equation}\label{eq:cohesive_energy}
    E_c = (E_{\mathrm{Cu}_n/surf} - E_{surf} - n\times E_{\mathrm{Cu_{FCC}}}/4) / n,
\end{equation}
respectively, where $E_{\mathrm{Cu}_n/surf}$ is the potential energy of a Cu nanoparticle on an Al$_2$O$_3$ surface, $E_{surf}$ is the potential energy of a pristine Al$_2$O$_3$ surface, $E_{\mathrm{Cu}_n}$ is the potential energy of the global minimum configuration of Cu$_n$ in the gas phase, and $E_\mathrm{Cu_{FCC}}$ is the total energy of an FCC Cu bulk unit cell that contains 4 atoms. The cohesive energy $E_c$ of a Cu nanoparticle in vacuo is computed from \textbf{Equation~\ref{eq:cohesive_energy}} with $E_{\mathrm{Cu}_n/surf}$ set equal to $E_{\mathrm{Cu}_n}$ and $E_{surf}$ set equal to zero.

\subsection{Molecular Dynamics}
We utilize MD to enrich the data set with structures at finite temperature and to investigate the supported nanoparticle dynamics.
All simulations are in the NVT ensemble controlled by a Langevin thermostat with a friction coefficient of 10 ps$^{-1}$ if not otherwise specified.
To explore configurations at finite temperature, MD sampling includes several active learning cycles, each of which includes NVT simulations at temperatures of 300 K, 600 K, 900 K, and 1200 K, respectively.  
From the trajectories, we select the most representative structures for further learning using two filters. 
The first is the uncertainty of the prediction of the model, a metric that has been used in many studies.\cite{Zhang2019PRM} 
The second filter is based on diversity.\cite{Bernstein2019npjComputMater}
For each structure, a vector is generated to represent its characteristics by adopting the SOAP descriptor\cite{Bartok2013PRB}, which is calculated as described in Ref. \cite{Himanen2020CPC,Laakso2023JCP}.
Subsequently, a similarity kernel is built from the SOAP vectors and a matrix decomposition algorithm\cite{Mahoney2009PNAS} is used to select the most representative structures.
The MD-based active learning strategy is detailed in \textbf{Section~S2.2}.

The mobility of Cu atoms in the supported nanoparticles is measured by their mean squared displacement MSD$_\mathrm{Cu}$, which is defined as
\begin{equation}\label{eq:msd}
    MSD(t;r) = {<\frac{1}{n}\sum_{i=1}^n \lVert r(t+\tau)-r(\tau)\rVert^2>}_{\tau},
\end{equation}
where $n$ is the number of particles (Cu atoms), $r(t)$ is the coordinate at timestep $t$, $\tau$ is the time origin, and $<\dotsb>_{\tau}$ denotes the average over all time origins with a lagtime of 0.1 ps.
The diffusion coefficient of COM, the center of mass of a nanoparticle, is calculated from
\begin{equation}\label{eq:diffusion_coefficient}
    D_\mathrm{COM}^d = \frac{1}{2d}\mathrm{lim}_{t\rightarrow\infty}\frac{\mathrm{d}}{\mathrm{d}t}\mathrm{MSD_{COM}^d},
\end{equation}
where $d$ is the spatial dimension, which is equal to 2 in the present work, where diffusion occurs on the surface X-Y plane, and MSD$_\mathrm{COM}^d$ is the MSD of the COM of a nanoparticle computed using 
\textbf{Equation~\ref{eq:msd}} and the number of particles $n$ becomes 1. 

To measure the dynamic MSI we use the bond correlation functions $C(t)$ defined by
\begin{equation}\label{eq:bond_correlation}
    C(t) = <\frac{<h_{ij}(t+\tau)h_{ij}(\tau)>_{ij}}{<h_{ij}(\tau)h_{ij}(\tau)>_{ij}}>_{\tau},
\end{equation}
where $h_{ij}=1$ if a bond exists between atoms $i$ and $j$ and $h_{ij}=0$ otherwise, $<\dotsb>_{ij}$ denotes the average over all the $ij$-bonds present when $t=\tau$, the time origin, and $<\dotsb>_{\tau}$ denotes the average over all time origins with a lagtime of 1 ps.
In this way, the Cu-Al and Cu-O bond correlation functions were studied. Here, Cu and Al were considered bonded if their distance was less than 3.0 \r{A}, while Cu and O were considered bonded if their distance was less than 2.2 \r{A}.
These two distances are slightly larger than the typical bond distances $d_{\mathrm{Cu}\text{-}\mathrm{Al}}\approx 2.6$ \r{A} of the bulk FCC CuAl and $d_{\mathrm{Cu}\text{-}\mathrm{O}}\approx 1.8$ \r{A} of the bulk crystalline Cu$_2$O.
Different cutoff values were tested, finding the same trends for the correlation functions (\textbf{Figure~S29}).

In the sintering simulations, we measure the minimum distance between the COMs of a pair of nanoparticles, which is given by  
\begin{equation}\label{eq:min_com_distance}
    d_{min}^\mathrm{COM} = \min_{1 \leq I\neq J \leq N} \lVert R_I^\mathrm{COM} - R_J^\mathrm{COM} \rVert,
\end{equation}
where $R_I^\mathrm{COM}$ is the COM of the $I$th nanoparticle and $N$ is the number of nanoparticles in the system.
Similarly, the minimum distance between two Cu atoms is defined by 
\begin{equation}\label{eq:min_copper_distance}
    d_{min}^\mathrm{Cu} = \min_{1 \leq I\neq J \leq N, i \in I, j \in J} \lVert r_i - r_j \rVert,
\end{equation}
where $r_i$ is the coordinate of the $i$th Cu atom belonging to the $I$th nanoparticle.

\begin{acknowledgement}
    The funding for this project was provided by Shell International Exploration and Production Inc., USA. The calculations were performed largely using the Princeton Research Computing resources at Princeton University. The authors also acknowledge the computing resources from the National Energy Research Scientific Computing Center (NERSC) operated under Contract No. DE-AC0205CH11231 using NERSC award ERCAP0021510. The software used in the present work has been developed in the Computational Chemical Science Center “Chemistry in Solution and at Interfaces (CSI)” funded by the USA Department of Energy under award DE-SC0019394. RC was partially supported by this grant, and JX benefited of scientific discussions with the members of CSI. The authors also thank Zheng Yu for fruitful discussions.
\end{acknowledgement}

\begin{suppinfo}
Supplementary Figures 1-34, Tables 1-5, notes, and references.
\end{suppinfo}

\bibliography{references}

\end{document}